\documentclass[11pt]{article}

\def\be{\begin{equation}}
\def\ee{\end{equation}}

\def\a{\alpha}
\def\b{\beta}
\def\t{\theta}
\def\tt{\tilde\theta}
\def\g{\gamma}
\def\={\ \ \ \ \Leftrightarrow \ \ \ \ }
\def\-{\longrightarrow}
\def\Pf{{\rm Pf}}

\usepackage{latexsym}
\usepackage{graphicx}

\begin{document}

\begin{flushright}
SU-4252-779\\
\today
\end{flushright}

\bigskip
\bigskip

\begin{center}

{\Large \bf A lattice study of the two-dimensional Wess Zumino model}

\bigskip
\bigskip

{\small Simon Catterall and Sergey Karamov\\
Physics Department, Syracuse University, Syracuse, NY 13244}

\footnotetext{Corresponding author: Simon Catterall, 
email: smc@physics.syr.edu}

\end{center}

\begin{abstract}

We present results from a numerical simulation   
of the two-dimen-\\
sional Euclidean Wess-Zumino model.
In the continuum the theory possesses $N=1$ supersymmetry.
The lattice model we employ was analyzed by Golterman and Petcher
in \cite{susy} where a perturbative proof was given 
that the continuum
supersymmetric Ward identities are recovered
without finite tuning in the limit of vanishing lattice
spacing. 
Our simulations demonstrate the existence of
important non-perturbative effects in finite volumes which modify
these conclusions. It appears that in certain regions of
parameter space the vacuum state can contain solitons corresponding
to field configurations which interpolate between different classical vacua.
In the background of these solitons supersymmetry is partially broken
and a light fermion mode is observed. At fixed
coupling the critical mass separating phases of broken and unbroken
supersymmetry appears to be
volume dependent. We discuss the implications of our
results for continuum supersymmetry breaking.

\end{abstract}

\newpage
\section{Introduction}

Supersymmetry has often been invoked as a necessary ingredient
for any particle physics theory which attempts to bridge the gap
between the scale of electroweak symmetry breaking and the much
larger scale associated to unification of the low energy gauge
interactions.
The basic idea is that while generic field theories
involving scalars are unstable to large radiative corrections which mix
scales, these 
radiative effects can be made much smaller if the scalar theory is 
embedded
inside some supersymmetric theory. The dynamical breaking of
supersymmetry through non-perturbative effects can then occur at
scales which are exponentially suppressed relative to the grand
unified scale. This symmetry breaking can, in turn, then trigger
electroweak breaking. 

Thus the non-perturbative structure of supersymmetric theories is
a subject of great interest. The only tool for a systematic investigation
of non-perturbative effects is the lattice and so a lot of
effort has gone into formulating lattice supersymmetric theories
\cite{general}. Typically it is difficult to write down lattice
actions which can be shown to flow to a supersymmetric
fixed point, without fine tuning, as the lattice spacing is reduced. 

The model we examine in this paper -- the two dimensional Wess-Zumino model
appears to provide an exception to this rule.  This theory
involves the interactions of scalars and fermion fields and exhibits
an $N=1$ supersymmetry in the continuum.
A version of this model defined 
on complex fields and
possessing $N=2$ supersymmetry was the subject of a recent numerical 
study
in \cite{wz} and 
was also examined in a variety of earlier papers \cite{wz2gen}.
The $N=2$ model actually possesses an exact
lattice supersymmetry which can be seen to result from its proximity to
a continuum topological field theory \cite{top}.

We have chosen to study a particular Euclidean
lattice formulation of the $N=1$ 
model
due to Golterman and Petcher \cite{susy}. The model
has also been studied using a Hamiltonian formulation in \cite{bec1} and
\cite{bec2}. Unlike the Hamiltonian formulations, the Euclidean
lattice theory does not retain {\it any} exact supersymmetry.
Nevertheless, Golterman
et al. prove that the
discrete analog of the continuum supersymmetric Ward identities are
satisfied exactly in the limit of vanishing lattice spacing without
the necessity of additional fine tuning. The proof is perturbative
and our goal in these simulations was to check whether the model
allows for supersymmetry breaking via non-perturbative
effects. We find that indeed the lattice model shows
evidence of supersymmetry breaking for small values of the lattice
mass parameter. Furthermore, this breaking is correlated to the
onset of field configurations which sample both the classical
vacua of the model. In this limit we 
also observe a light fermion state which we
speculate may play the role of a Goldstino
associated with spontaneous supersymmetry breaking.

We have developed and tested
a fourier accelerated version of the so-called R-algorithm 
\cite{Ralg} to handle the fermionic integrations. 
For details of this Fourier acceleration technique in the
context of the HMC algorithm we refer the reader to \cite{FFT}.
We have employed an exact algorithm to calculate the
sign of the Pfaffian resulting from the integration over the
fermion fields.
These issues are discussed in detail in
section 2. We present our evidence for
symmetry breaking together with numerical results on the spectrum
and Ward identities in
section 3. 
In section 4 we summarize our findings and discuss their implications for
supersymmetry breaking in the continuum in finite and infinite volume.

\newpage
\section{Lattice model}

We consider the on-shell two-dimensional Wess-Zumino model
represented by the following continuum action in Euclidean space
\cite{susy}:
\be
S_0= \int d^2x \ \frac12 \ 
[(\partial_\mu\phi)^2+\bar\psi(\partial\!\!\!/+P'(\phi))\psi+P^2(\phi)]
\label{contact}
\ee
where $\phi$ and $\psi$ are a real scalar field and a two component 
Majorana spinor respectively. The construction of 
Euclidean Majorana spinors
is described by Nicolai in \cite{nico}.
The expression $Q(\phi)=\partial\!\!\!/+P'(\phi)$ will be referred to 
as the fermion matrix.
The potential $P(\phi)$ we consider (actually the derivative of
the superpotential)  takes the following 
form depending on a mass $M$ and a coupling constant $G$:
$$
P(\phi)=
\left\{ 
\begin{array}{l}
M\phi\\
G\phi^2-M^2/4G
\end{array} 
\right.
\left. \begin{array}{l}
,\ \ G=0\\
,\ \ G\ne0
\end{array} \right.
$$
Notice that this potential is slightly different from the
one considered in \cite{susy} but may be derived from it by a simple
shift in the scalar field. It has the advantage that
the total action now only depends on $M^2$ which
allows us to restrict our simulations to positive $M$.
Notice also that the interacting theory has two classical vacua
at $\phi=\pm M/2G$.
The action (\ref{contact}) is invariant under the following 
supersymmetry transformation:
$$
\delta \phi=\bar \varepsilon \psi,\ \ \ \ 
\delta \psi=(\partial\!\!\!/\phi-P(\phi)) \varepsilon 
$$
The simplest supersymmetric Ward identity following from this
invariance takes the form
\be
<\psi_x \bar \psi_y>+<[\partial\!\!\!/\phi-P(\phi)]_x\phi_y>=0
\label{ward}
\ee
Integrating out fermion variables in the path integral 
leads to the following form of the partition function \cite{Montvay}
(see the Pfaffian definition (\ref{Pf}) in appendix):
$$
Z=\int D\phi D\psi e^{-S_0}=
\int D\phi \ {\rm Sign}[{\rm Pf}(CQ)] 
\ e^{{\rm tr}[\ln(Q^TQ)]/4-S_b}
$$
where C is a Euclidean representation of the charge conjugation 
matrix and $S_b$ stands for the bosonic part of the action:
$$
S_b=\int d^2x \ \frac12 \ [(\partial_\mu\phi)^2+P^2(\phi)]
$$
In practice we simulate the system without regard to the
sign of the Pfaffian using the following action $S$
\be
S=-\frac14 {\rm tr}[\ln(Q^TQ)] + 
\int d^2x \ \frac12 \ [(\partial_\mu\phi)^2+P^2(\phi)]
\label{redef}
\ee
The expectation values of physical observables are then obtained
by reweighting with the measured sign of the Pfaffian in the usual
manner
\be
<O>_{S_0}=\frac{<O\ {\rm Sign}[{\rm Pf}(CQ)]>_S}
{<{\rm Sign}[{\rm Pf}(CQ)]>_S}
\label{aver}
\ee

We now turn to the lattice model. First, we replace the continuum
derivative operator by the symmetric difference matrix 
$D^\mu_{\bf r r^\prime}$
where the latter is defined as:
$$
D^\mu_{\bf r r^\prime}=
\frac12 [\delta_{\bf r+e_\mu, r^\prime}-
\delta_{\bf r-e_\mu, r^\prime}]
$$
where ${\bf r}$, ${\bf r^\prime}$ are two-dimensional vectors 
enumerating the lattice sites and ${\bf e_\mu}$ is a unit vector 
in the $\mu$-direction $\mu=1,2$. In terms of this difference operator
the fermion matrix on the lattice can be represented as:
$$
Q \equiv Q_{\bf r r^\prime}^{\alpha \beta}=
\gamma^\mu_{\alpha \beta} D^\mu_{\bf r r^\prime}
+ \delta_{\alpha \beta} P^\prime_{\bf r r^\prime}
$$
where $\alpha$, $\beta$ are spinor indices.
We have employed the following representations of the
Dirac matrices\[
\begin{tabular}{ccc}
$\gamma_1=\left(\begin{tabular}{cc}
1&0\\0&-1\end{tabular}\right)$&
$\gamma_2=\left(\begin{tabular}{cc}
0&-1\\-1&0\end{tabular}\right)$
\end{tabular}
\]
The matrix $C$ is given explicitly as
$$
C=\left(\begin{tabular}{cc}
0&-1\\1&0\end{tabular}\right)
$$
It is convenient to define an operator 
$\Box^n_{\bf r r^\prime}$:
$$
\Box^n_{\bf r r^\prime}=\frac12 \sum_\mu[
\delta_{{\bf r}+n{\bf e_\mu}, {\bf r^\prime}}+
\delta_{{\bf r}-n{\bf e_\mu}, {\bf r^\prime}}-
2\delta_{\bf r, r^\prime}]
$$
In particular
$$
\Box^2_{\bf r r^\prime}=(D^\mu D^\mu)_{\bf r r^\prime}
$$
In terms of the operator $\Box^n_{\bf r r^\prime}$ 
the lattice potential and its derivative
can be represented as follows:
$$
P_{\bf r}=
\left\{ 
\begin{array}{l}
m\phi_{\bf r}-\Box^1_{\bf r r^\prime} \phi_{\bf r^\prime}/2\\
g\phi_{\bf r}^2-m^2/4g-\Box^1_{\bf r r^\prime} \phi_{\bf r^\prime}/2
\end{array} 
\right.
\left. \begin{array}{l}
,\ \ g=0\\
,\ \ g\not=0
\end{array} \right.
$$
$$
P^\prime_{\bf r r^\prime} \equiv
\frac{\partial P_{\bf r}}{\partial \phi_{\bf r^\prime}} =
\left\{ 
\begin{array}{l}
m\delta_{\bf r r^\prime}-\Box^1_{\bf r r^\prime}/2\\
2g\phi_{\bf r}\delta_{\bf r r^\prime}-\Box^1_{\bf r r^\prime}/2
\end{array} 
\right.
\left. \begin{array}{l}
,\ \ g=0\\
,\ \ g\not=0
\end{array} \right.
$$
where the term with $\Box^1_{\bf r r^\prime}$ is the Wilson mass operator, 
which serves to eliminate 
problems due to doubling of the lattice fermion modes and vanishes 
in the continuum limit. The dimensionless lattice couplings $g$ and $m$
are related to their continuum counterparts through the relations
$g=Ga$ and $m=Ma$ with $a$ the lattice spacing.

Finally the lattice representation of the continuum action 
(\ref{redef}) can be viewed as the sum of the following 
boson and fermion components:
$$
S_b=\frac 12 \lbrace - \phi_{\bf r} 
\Box^2_{\bf r r^\prime} \phi_{\bf r^\prime}  + 
P_{\bf r} P_{\bf r} \rbrace
$$
$$
S_f = - \frac14 {\rm tr}[\ln(Q^TQ)] \equiv 
- \frac 14 [\ln(Q^TQ)]_{\bf r r}^{\alpha \alpha}
$$

\newpage
\section{Simulation Details}

To simulate the system (\ref{redef})
we use an importance sampling technique based on a classical
evolution of the fields in some auxiliary time.
To implement this it is necessary to introduce
a Hamiltonian 
$$
H=\frac12 p_{\bf r}p_{\bf r} + S
$$
associated with this auxiliary time variable $t$ and corresponding
momentum field $p$ conjugate to the field $\phi$. On integration
over the auxiliary momentum $p$ it is trivial to show
that the classical partition function associated to $H$ reproduces the
quantum partition function associated with the original
action $S$.
The advantage of this Hamiltonian formulation is that it admits 
a classical dynamics, which can be used to generate global moves
of the field $\phi$.

We evolve the system governed by $H$ according to a finite time
step leapfrog algorithm in the usual manner
$$
\left. \begin{array}{l}
\phi_{t+\delta t} = \phi_t + p_t \delta t + F_t (\delta t)^2/2\\
p_{t+\delta t} = p_t + (F_t+F_{t+\delta t})\delta t/2
\end{array} \right.
$$
where $F$ is the force associated with the classical evolution.
The ergodicity of the simulation is provided by periodically
drawing new momenta $p$ from a Gaussian distribution.
In order to decrease the autocorrelation time associated with this
dynamics we have utilized acceleration techniques similar to those
explored in \cite{FFT}.
Specifically, the discrete time update of the fields
corresponding to the Hamiltonian evolution is carried out in momentum 
space
with a momentum dependent time step which is tuned so as to evolve 
low momentum components of the field more rapidly than high momentum
components. Specifically we used a time step of the form
$$
\delta t({\bf n})=
\epsilon \frac{m_{acc}+4}
{\sqrt{\sum_{\mu=1}^2\sin^2{\frac{2\pi n_\mu}{L}}+\left(m_{acc}+
2\sum_{\mu=1}^2\sin^2{\frac{\pi n_\mu}{L}}\right)^2}}
$$
where the lattice momenta $\bf n$ are integer vectors with
components ranging from $0\to L-1$ for an $L\times L$ lattice.
The parameter $m_{acc}$ is typically set to the input lattice
mass which is close to optimal in these simulations.
 
The total force can be represented as a sum of 
boson and fermion contributions:
$$
F_{\bf r}=F^b_{\bf r}+F^f_{\bf r}
$$
The evaluation of the boson force is straightforward:
$$
F^b_{\bf r}=-\frac{\partial S_b}{\partial b_{\bf r}}=
\Box^2_{\bf r r^\prime} \phi_{\rm r^\prime}
- P_{\bf r^\prime} P^\prime_{\bf r^\prime r}
$$
In order to evaluate the fermion force we first evaluate the 
following expression involving the fermion matrix:
$$
\frac{\partial (Q^TQ)_{\bf r r^\prime}^{\alpha \beta}}
{\partial b_{\bf s}}=
\frac {\partial^2 P_{\bf s^\prime}}
{\partial b_{\bf r} \partial b_{\bf s}}
Q_{\bf s^\prime r^\prime}^{\alpha \beta} 
+
Q_{\bf s^\prime r}^{\beta \alpha} 
\frac {\partial^2 P_{\bf s^\prime}}
{\partial b_{\bf r^\prime} \partial b_{\bf s}}=
2g(
\delta_{\bf r s}
Q_{\bf r r^\prime}^{\alpha \beta} 
+
Q_{\bf r^\prime r}^{\beta \alpha} 
\delta_{\bf r^\prime s}
)
$$
The fermion force then is:
$$
F^f_{\bf s}=-\frac{\partial S_f}{\partial b_{\bf s}}=
\frac 14 [(Q^TQ)^{-1}]^{\alpha \beta}_{\bf r r^\prime}
\frac{\partial (Q^TQ)^{\beta \alpha}_{\bf r^\prime r}}
{\partial b_{\bf s}}=
g[(Q^TQ)^{-1}]^{\alpha \beta}_{\bf s r}
Q^{\alpha \beta}_{\bf s r}
$$

The computation of the fermion force appears to be problematic as it 
appears to require
the repeated inversion of the fermion matrix
which is prohibitively expensive.
In order to resolve this problem we use the so-called R-algorithm 
\cite{Ralg}.
The algorithm proceeds by replacing the exact inverse matrix
$(Q^TQ)^{-1}$ by a stochastic estimator
given by the following expression
\be
[(Q^TQ)^{-1}]^{\alpha \beta}_{\bf r r^\prime} \approx
<X^\alpha_{\bf r} X^\beta_{\bf r^\prime}>_N
\label{approx}
\ee
where the vector $X$ is defined through a random Gaussian 
vector $R_g$ as:
$$
QX=R_g
$$
and the averaging in (\ref{approx}) is accomplished 
over $N$ different random noise vectors $R_g$.

\begin{figure}[htb]
\begin{center}
\includegraphics[width=11cm]{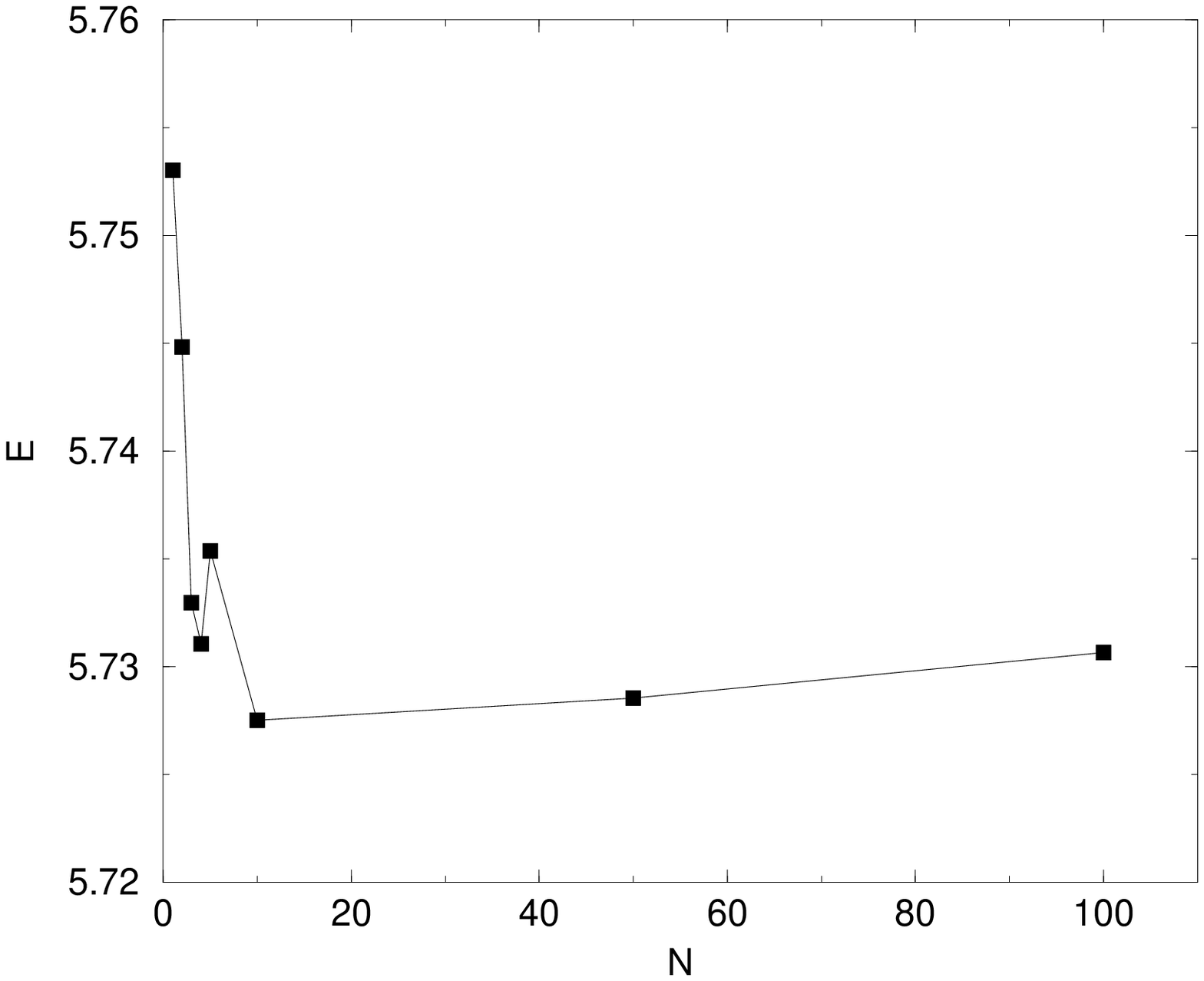}
\end{center}
\caption{\label{MM} Algorithm efficiency as 
a function of the number of noise vectors.}
\end{figure}

The larger the number of noise vectors used $N$ the more 
accurate is the evaluation of the inverted matrix in (\ref{approx}),
but the longer computational time the evaluation takes.
It is clear that the optimal value of $N$ is given by that
which minimizes the error in the inverse matrix for fixed computational
time $T$.
Defining the norm of a matrix $\|A\|$ as
$$
\|A\|=\sqrt{\sum_{ij}A^2_{ij}}
$$
The relative error is then
$$
\frac{\delta \|A\|}{\|A\|}=
\sqrt{\frac NT} \left\{ \frac {\delta \|A\|}{\|A\|} \right\}_N
$$
where $\{\delta \|A\|/\|A\|\}_N$ is the relative error 
produced by a single application of an R-algorithm
with averaging over $N$ noise vectors. 
Hence the relative error obtained over time $T$ can be 
characterized by the algorithm efficiency $E$ which we define as
$$
E=\sqrt N \left\{ \frac {\delta \|A\|}{\|A\|} \right\}_N
$$
Our tests showed that  
this algorithm efficiency does not depend strongly on 
the choice of $N$ (Figure \ref{MM}). Furthermore, we
monitored the average bosonic action $<S_b>$ and observed no
systematic drift with $N$.
Consequently we chose $N=1$ in all our runs.
In this limit the corresponding fermion force term yields 
$$
F^f_{\bf s}\approx
\frac 14 X^\alpha_{\bf r} X^\beta_{\bf r^\prime}
\frac{\partial (Q^TQ)^{\beta \alpha}_{\bf r^\prime r}}
{\partial b_{\bf s}}=
\frac g2 Q^{\alpha \beta}_{\bf s r} 
(X^\alpha_{\bf s} X^\beta_{\bf r} +
X^\beta_{\bf s} X^\alpha_{\bf r})
$$

Finally let us turn to the issue of the sign of the
Pfaffian which results from integrating out the Majorana
fields. As we have stressed the simulation
action discussed above utilizes the absolute value of this Pfaffian and
observables must be re-weighted by the sign of the
Pfaffian in order to compute physical
expectation  values. We have chosen to use an exact algorithm to compute
this sign. Since we are in two dimensions and need only do this
when making measurements this turns out to be quite
manageable in a practical sense. 
Our procedure was outlined in \cite{campos} and for completeness we
list the proof and details of the algorithm in an appendix. In essence
the original antisymmetric matrix can be transformed to a special block
diagonal form
via a similarity transformation built from a triangular matrix. The
determinant of the latter can be shown to yield the Pfaffian. We then
fold
the sign of the Pfaffian in with measurements of observables 
according to (\ref{aver}). This reweighting 
procedure is an effective way to measure expectation
values of a variety of observables. However, in certain conditions
this technique may fail. The following arguments highlight the
problems that may be encountered in this type of situation.

Let $N_+$ and $N_-$ be the numbers of configurations with positive
and negative values of $\Pf(CQ)$ obtained from the simulation of 
the system 
(\ref{redef}). Then the average value $<O>_{S_0}$ 
of any physical observable $O$ in the system (\ref{contact})
can be {\it evaluated} using (\ref{aver}) as:
\be
<O>=\frac{O_+N_+-O_-N_-}{N_+-N_-}
\label{averNN}
\ee
where $O_\pm$ are average values of $O$  obtained in
configurations subsets with positive and negative $\Pf(CQ)$. 
This averaging procedure reveals two statistical problems. 
The first problem is that if 
$N_+ \approx N_-$ (that is the probabilities to find the system
with either sign of the Pfaffian are approximately the same)
then the error of the evaluation (\ref{averNN}) experiences 
an amplification by a large factor $(N_++N_-)/(N_+-N_-)$.
In this case it is possible for the error to swamp the signal 
in the measured value $<O>$. Although
acquiring more measurements will decrease the fluctuations 
it might not solve the amplification problem if
\be
\lim_{N_++N_-\rightarrow \infty}\frac{N_+}{N_-} \sim 1
\label{N/N}
\ee 

A second problem is that the expression (\ref{averNN})
provides a good evaluation of $<O>_{S_0}$ only if $<O>$ 
is uniquely defined in the limit 
$N_++N_-\rightarrow \infty$, which is not necessarily
the case. If (\ref{N/N}) takes place
then $<O>$ is well defined only if 
$<O_+>=<O_->$, which is not guaranteed to be true.

In practice we find that many of our observables suffer large and
difficult to quantify errors for small values of the lattice mass
where we observe oscillations in the sign of the Pfaffian. This 
precludes making strong quantitative statements in that region.

\newpage
\section{Results}

\begin{figure}[htb]
\begin{center}
\includegraphics[width=11cm]{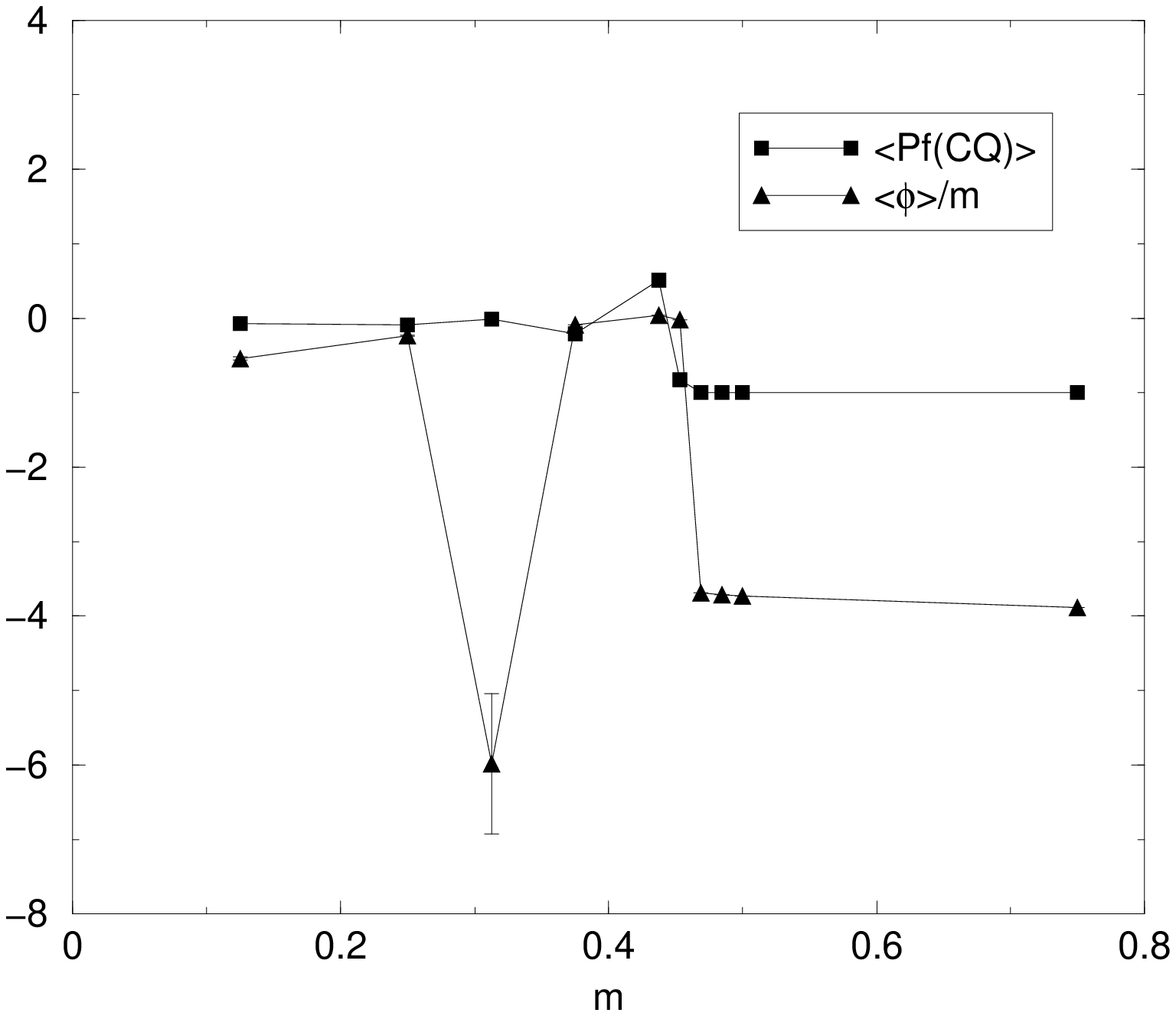}
\end{center}
\caption{\label{8Pf} Average field
and average $\Pf (CQ)$ for L=8. The field values are rescaled 
for $m<0.46$ by a factor of 1/200.}
\end{figure}

\begin{figure}[htb]
\begin{center}
\includegraphics[width=11cm]{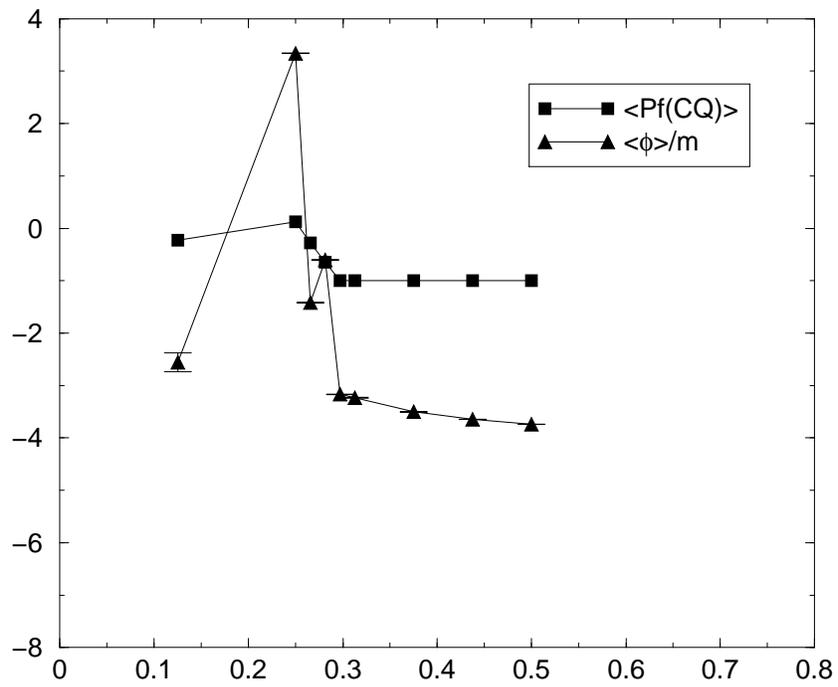}
\end{center}
\caption{\label{16Pf} Average field
and average $\Pf (CQ)$ for L=16. The field values are rescaled 
for $m<0.29$ by a factor of 1/8.}
\end{figure}

\begin{figure}[htb]
\begin{center}
\includegraphics[width=11cm]{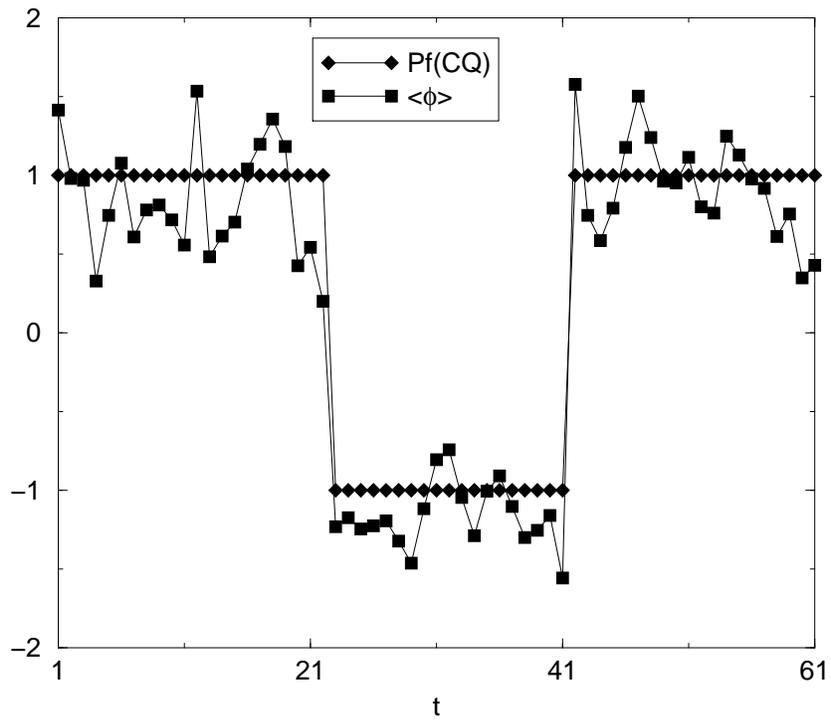}
\end{center}
\caption{\label{PfB} 
Evolution of $\Pf(CQ)$ and $<\phi>$ in auxiliary time t
for L=8, m=0.125.}
\end{figure}

We obtained data for lattice sizes
$L=8$ and $L=16$ for a fixed lattice coupling $g=0.125$ while
varying the lattice bare mass $m$.
The classical vacua of the lattice theory correspond to
vanishing fermion field and boson field $\phi=\pm m/2g$.
For large $m$, field configurations which interpolate
between these two vacua are associated with large values of the
action and are hence expected to be highly suppressed. We thus
expect the boson
field to be confined in the neighborhood of one of the
classical vacua for sufficiently large mass. In the continuum the action is
invariant under $\phi\to -\phi$ implying that these two
vacuum states are equivalent. This is
no longer true
on the lattice due to the presence of the Wilson term (actually the
sign of the Pfaffian may also change under this symmetry). 
Indeed our simulations reveal that only the state with
$<\phi>\sim -m/2g$ survives at large $m$. 
As $m$ decreases we
expect
that tunneling to the other vacuum state may occur and this is
indeed seen in our simulations.
Figures \ref{8Pf} and \ref{16Pf}
show plots of $<\Pf (CQ)>$ and $<\phi>/m$ 
versus $m$ for $L=8,16$. Below some
critical $m=m_c(L)$ the sign of the Pfaffian 
which is negative at large mass $m$ starts to fluctuate.
Additionally, in this region we can see that 
the average field $<\phi>/m$
also undergoes large fluctuations  
which are the direct result of the Pfaffian sign changes.
Indeed, at small mass we observe that for
each configuration in our ensemble the sign of the Pfaffian is
very accurately correlated with the sign of the mean boson field.
Figure \ref{PfB} shows a time series of both quantities
at $m=0.125$ and $L=8$ which illustrates very well this behavior. 
Actually it is easy to see why this is so. Imagine expanding
the Pfaffian 
as a power series in the boson field $\phi$. For sufficiently
small $m$ we expect that only the leading term is important and
by translational symmetry this can depend only on the field
summed over all lattice sites. 
$$
\Pf(CQ)\sim \sum_{\bf r} \phi_{\bf r} 
+ O \left( \sum_{\bf r r'}\!' \phi_{\bf r} \phi_{\bf r'} \right)
$$ 
As we discussed in the previous section this sign oscillation 
renders accurate measurements of $<\phi>$ and its error very
difficult in this region.

\begin{figure}[htb]
\begin{center}
\includegraphics[width=11cm]{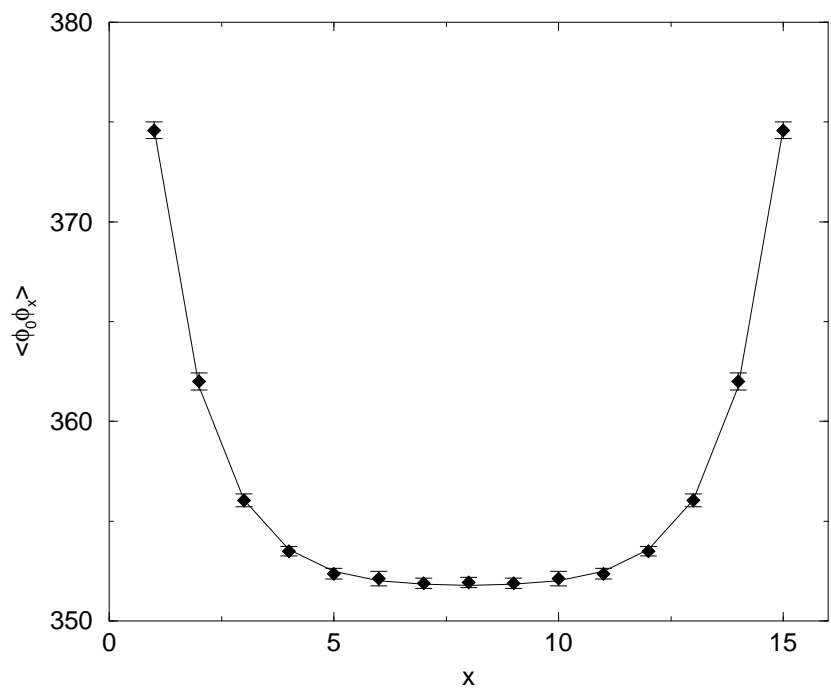}
\end{center}
\caption{\label{Bcorr} 
Bosonic correlation function for L=16, m=0.5.}
\end{figure}

\begin{figure}[htb]
\begin{center}
\includegraphics[width=11cm]{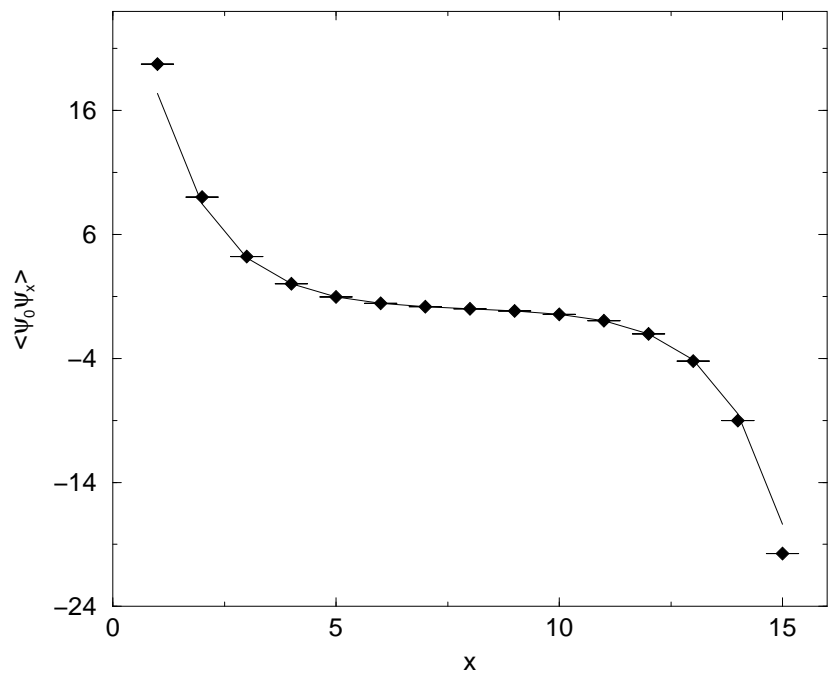}
\end{center}
\caption{\label{Fcorr} 
Fermionic correlation function for L=16, m=0.5.}
\end{figure}

\begin{figure}[htb]
\begin{center}
\includegraphics[width=11cm]{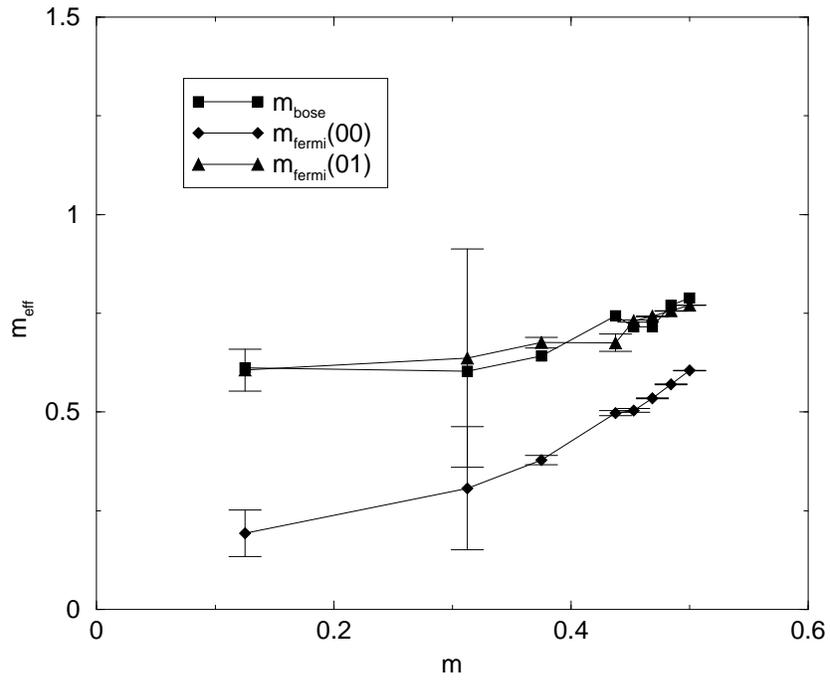}
\end{center}
\caption{\label{8mass} 
Mass gaps for L=8.
Mass gaps from bosonic correlators are shown for $m<0.46$
without error bars}
\end{figure}

\begin{figure}[htb]
\begin{center}
\includegraphics[width=11cm]{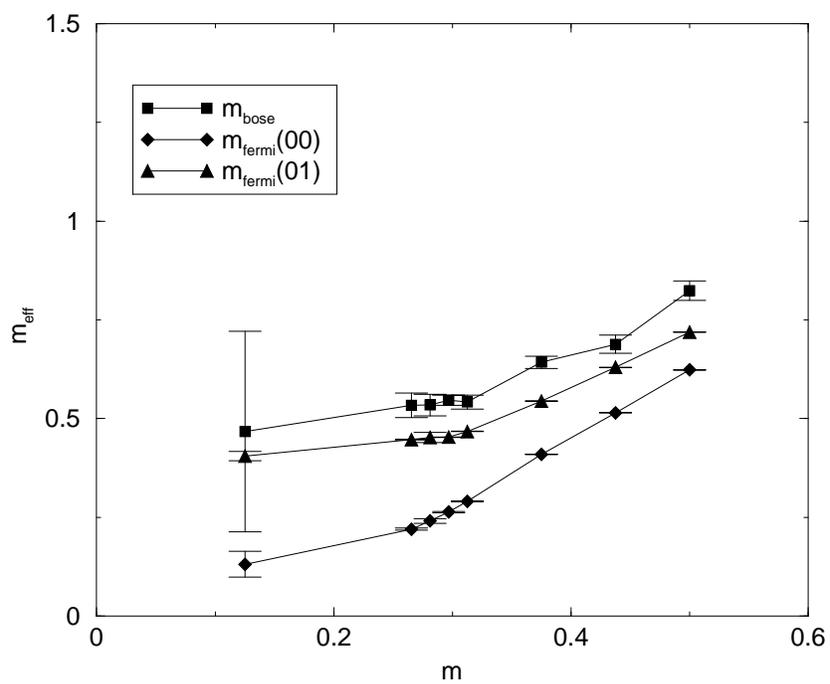}
\end{center}
\caption{\label{16mass} 
Mass gaps for L=16.}
\end{figure}

We have also measured the (zero momentum) boson and fermion 
correlation functions over the same range of lattice bare masses. 
Figures \ref{Bcorr} and \ref{Fcorr}
show typical bosonic and fermionic two point functions computed
on ensembles corresponding to 
$L=16$ with $m=0.5$. These are fitted by
hyperbolic $\cosh$ and mixed hyperbolic $\sinh$ and $\cosh$ functions
to extract the corresponding boson
and fermion masses. These (lattice) masses  are
shown in figures \ref{8mass} and \ref{16mass} for $L=8$ and $L=16$ 
respectively. The statistical errors we show neglect the effects
of correlation between observables at different timeslices.
Consider the data for $L=8$.
Notice that the boson and fermion masses 
are equal within statistical
errors at large bare input mass but deviate substantially at small
mass - the diagonal spinor components of the
fermion correlator being dominated by a light state. 
Contrast this will the off-diagonal components of the fermion
correlator for small bare mass which 
yield a much heavier mass degenerate
with the boson mass within statistical errors. A light fermion state
is also visible in the $L=16$ data at small mass. 
The mass of this light fermionic state appears to decrease with the bare input
lattice mass $m$. It is tempting 
to conclude from these observations that for small enough mass
supersymmetry breaks as a result of mixing between the
two classical vacua -- this
being signaled by the appearance of a Goldstino.

Another line of evidence in favor of this derives from the
partition function itself. On a finite lattice equipped with
periodic boundary conditions, such as employed
in our simulations, the partition function can be
thought of as yielding a representation of the
Witten index. Vanishing Witten index is a necessary condition
for supersymmetry breaking. But $Z$ can also be related to
the expectation value of the sign of the Pfaffian in our
simulation ensemble
$$
Z_{S_0}=<{\rm Sign}\left(\Pf (CQ)\right)>_S
$$
Thus we see that a vanishing partition function 
would require equal numbers of positive and negative sign Pfaffians
in our ensemble. Table 1. shows the numbers of
positive $N_+$ and negative $N_-$ Pfaffians for three different runs at
the same parameter values $L=8$ and $m=0.25$ each containing $100,000$
measurements.
\begin{table}
\begin{center}
\begin{tabular}{|c|c|}
\hline
$N_+$ & $N_-$\\ \hline
40968 & 59032\\ \hline
43814 & 56186\\ \hline
52252 & 47748\\
\hline
\end{tabular}
\caption{Numbers of positive and negative Pfaffians for $L=8$ and $m=0.25$}
\end{center}
\end{table}
While the relative errors of on the order of ten percent it should be
clear that the data are consistent with a vanishing Witten index
implying a non-zero vacuum energy.

\begin{figure}[htb]
\begin{center}
\includegraphics[width=11cm]{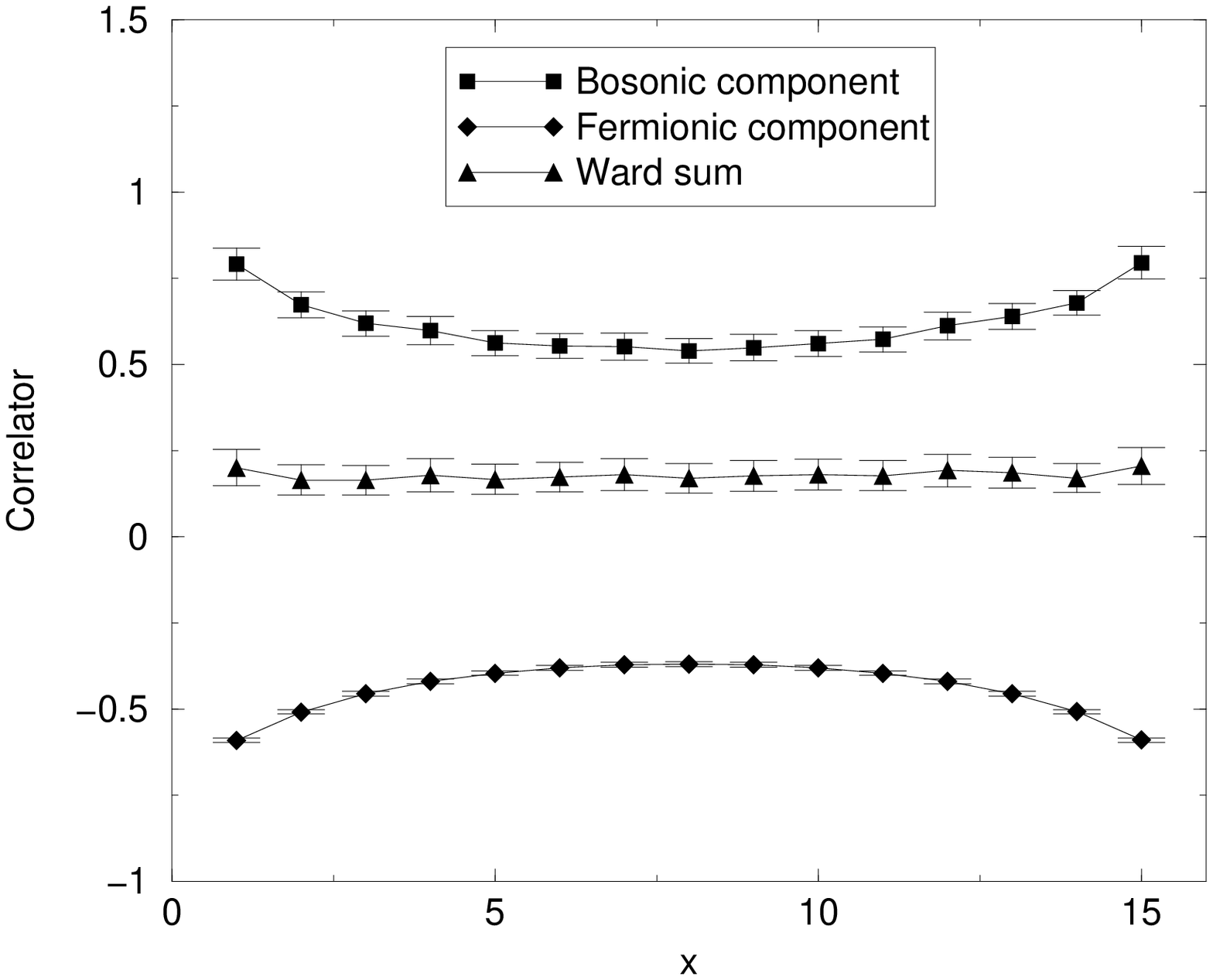}
\end{center}
\caption{\label{d1ward} 
Contributions to the diagonal components of 
Ward identity for m=0.125.}
\end{figure}

\begin{figure}[htb]
\begin{center}
\includegraphics[width=11cm]{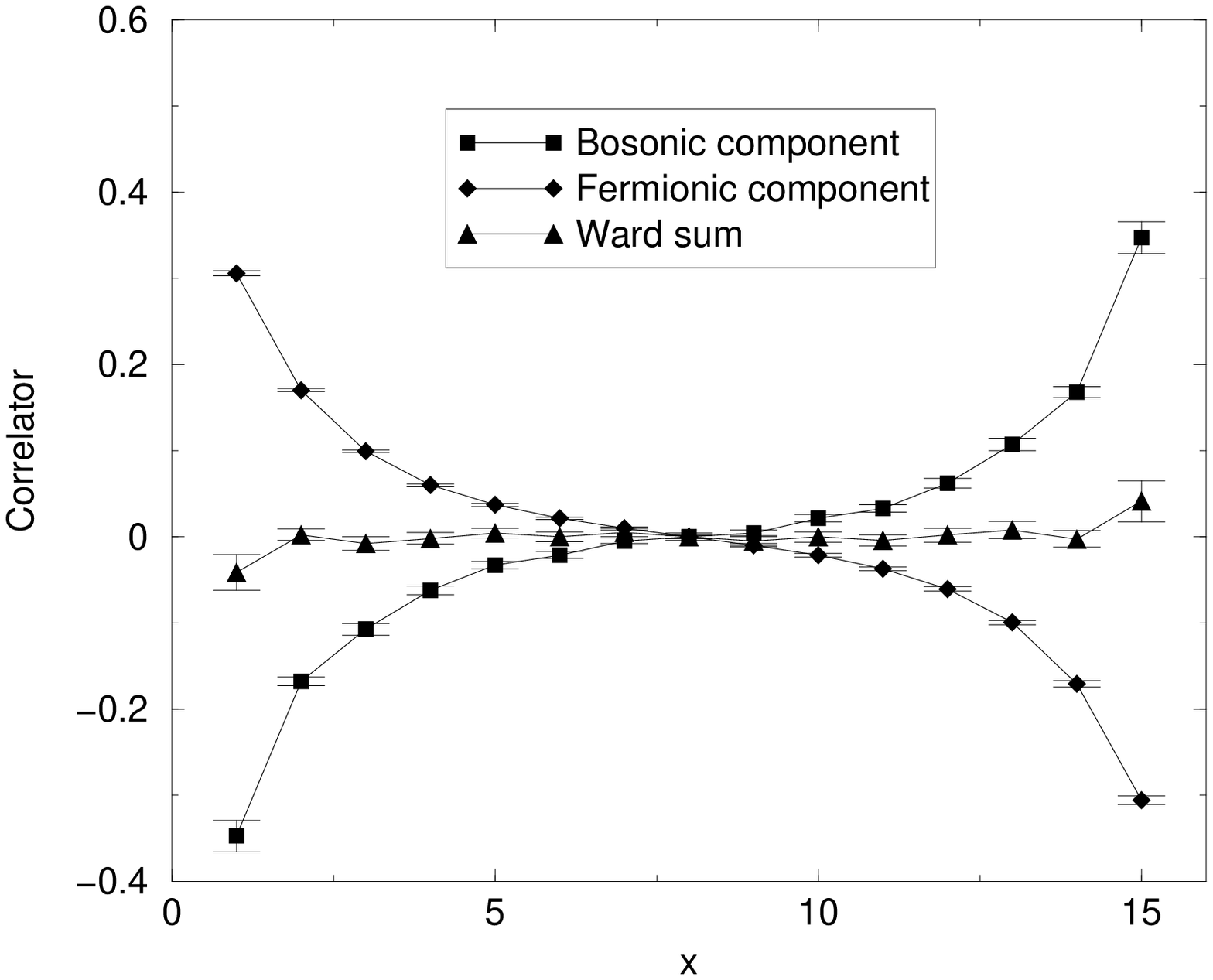}
\end{center}
\caption{\label{od1ward} 
Contributions to the off-diagonal components of 
Ward identity for m=0.125.}
\end{figure}

\begin{figure}[htb]
\begin{center}
\includegraphics[width=11cm]{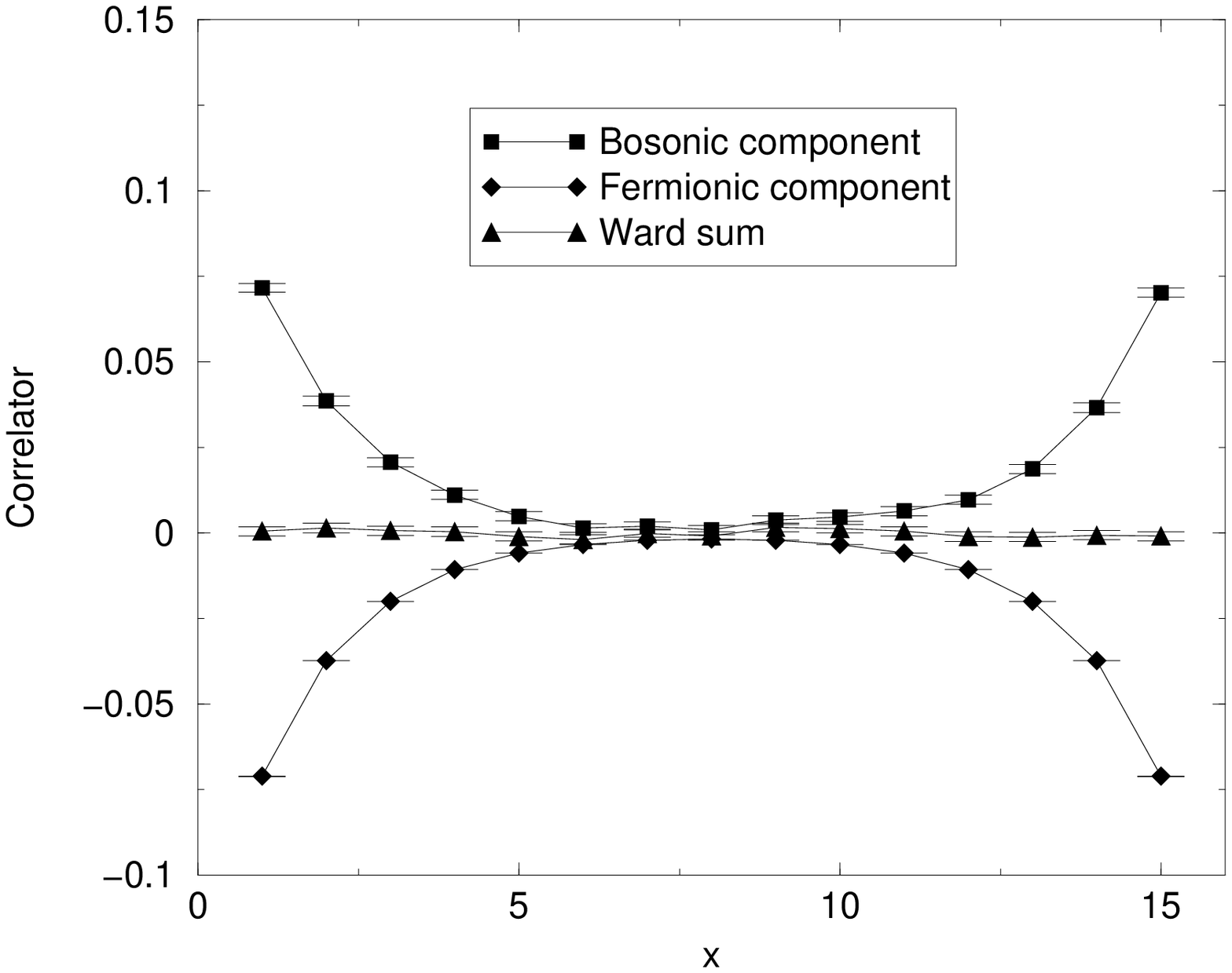}
\end{center}
\caption{\label{d4ward} 
Contributions to the diagonal components of 
Ward identity for m=0.5.}
\end{figure}

\begin{figure}[htb]
\begin{center}
\includegraphics[width=11cm]{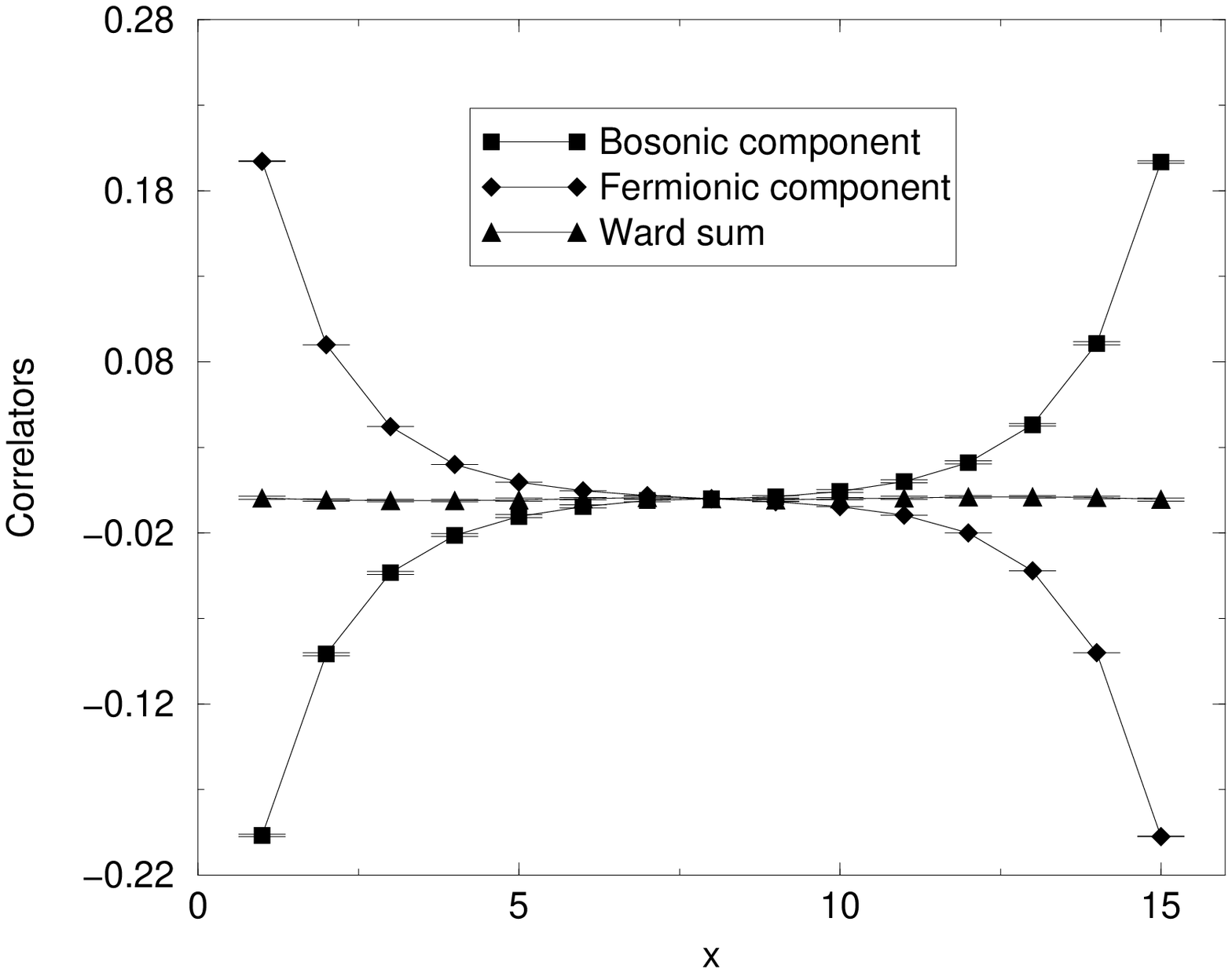}
\end{center}
\caption{\label{od4ward} 
Contributions to the off-diagonal components of 
Ward identity for m=0.5.}
\end{figure}

To investigate this symmetry breaking further we have looked at the simplest 
supersymmetric
Ward identity involving two point functions (\ref{ward}).
Figures \ref{d1ward} to \ref{od4ward} show the bosonic 
and fermionic diagonal and off-diagonal spinor contributions to this
Ward identity together with their sum for two different values of
the bare mass $m=0.125$ and $m=0.5$ on a lattice with
$L=16$. Clearly for large mass this
relation is satisfied within errors for all spinor
channels but it is clearly violated at 
small mass for the channel involving the diagonal spinor correlations.
The latter channel is precisely the one in which the light
fermion was seen and support the idea that breaking of supersymmetry
is associated with the appearance of a Goldstino. Notice that the
off-diagonal components of the Ward identity are {\it still} accurately
satisfied even at small mass. We will argue in the next section that
this is exactly what we might expect for a partial breaking of
supersymmetry associated with the appearance of a finite volume vacuum state 
composed of solitons.
\newpage
\section{Conclusions}

We have studied a lattice regularized version of the two-dimensional
Wess Zumino model which possesses $N=1$ supersymmetry in the 
naive continuum
limit. This model was first analyzed in \cite{susy} where it
was shown perturbatively that the supersymmetric Ward identities are recovered
without finite tuning in the limit of vanishing lattice
spacing. The goal of our simulations was to check these conclusions
at the non-perturbative level and to specifically to address
the important issue of supersymmetry breaking.
We have considered the model for fixed lattice coupling
$g=0.125$ and varying lattice mass $m$ for two
lattice sizes $L=8$ and $L=16$. For large $m$ 
our results favor a supersymmetric phase in which boson states
pair with equal mass fermion states and the supersymmetric
Ward identities are satisfied. In this region of
parameter space corresponding to $\phi\sim -m/2g$ 
the boson field suffers small fluctuations
around a single vacuum state.

As the mass is lowered however this picture changes and below some
critical mass $m_c(L)$ we see configurations in which the mean boson
field varies in sign
corresponding to tunneling between different vacua in auxiliary
time. The appearance of states which interpolate between
different perturbative vacua is of course entirely
a non-perturbative effect. 
Associated with these tunneling states we see oscillations in the Pfaffian
of the fermion operator and the appearance of a light fermion visible in the 
diagonal components of the fermion correlator. In such a phase
it appears that supersymmetry is at least partially broken.

It is possible to get some further understanding of this phenomenon
within the context of the semi-classical approximation. 
Consider first the continuum model. It is clear
that {\it in a finite volume} corresponding to a
box of size $L_{\rm phys}$, in addition to the
supersymmetric vacua $\phi=\pm M/2G$, there are additional local
minima of the action (\ref{contact}) corresponding to domain wall
solutions which interpolate between these vacua.
\begin{eqnarray*}
\phi(x)&\to&\Lambda,\;\;x\to\infty\\
\phi(x)&\to& -\Lambda,\;\;x\to -\infty
\end{eqnarray*}
where $\Lambda=M/2G$ and $x$ corresponds to one of the coordinate directions. 
Indeed, in the continuum, these solutions take the form
$$
\phi(x)=\Lambda\tanh(M(x-x_0)/2)
$$
While the mass of such a soliton state is non-zero it is possible to
show that it is nevertheless annihilated by a single component of
the Majorana supercharge and hence such a state preserves one half of
the original supersymmetry \cite{olive}. This is the origin of
our observation that certain components of the Ward identities appear
to be satisfied at all values of the parameters.
The action of such a soliton is easily evaluated
$S_{\rm DW}=\frac{4}{3}GL_{\rm phys}\Lambda^3$ and being proportional
to the integral of a total derivative term is topological in
character.
The corresponding free energy associated with such domain
wall solutions then varies as
$$
F_{\rm DW}\sim -\ln{L_{\rm phys}} +S_{DW}
$$
where the logarithmic variation with volume arises from the
number of ways the domain wall can be introduced into the finite
volume.
These arguments lead one to conclude that these 
non-supersymmetric vacua will dominate
over the supersymmetric vacua if
\be
\frac{S_{\rm DW}}{L_{phys}}=
\frac{M^3}{6G^2}<\left(\frac{\ln {L_{\rm phys}}}{L_{\rm phys}}\right)
\label{SDW}
\ee
At fixed $G$ this result is in qualitative agreement with
our lattice results since it predicts a critical mass $M_C(L_{\rm phys})$
below which supersymmetry would be broken.
Translating this result naively into lattice variables leads to
the prediction that $m_C\sim 0.3$ for $L=8$ and $g=0.125$ which is
quite close to the continuum estimate $M_Ca=0.46$ for $L_{\rm phys}=8a$.
According to our observations this critical mass 
shifts to smaller values as the lattice size 
increases which is also in agreement with these analytic arguments.
Furthermore, in the vicinity of such a domain wall the
fermion is approximately massless and so can play the
role of a Goldstino associated with
supersymmetry breaking. Notice that these arguments {\it rely} on
the constraint of {\it finite} volume. The action of such a soliton
is unbounded in infinite volume and hence we would naively expect solitons to
be completely suppressed in such a limit.

Of course we would like to know whether this finite volume supersymmetry
breaking scenario persists in the continuum limit. 
In general, in {\it finite} physical volume $V=L_{\rm phy}^2$, 
the continuum limit $a=1/L_{\rm phy} \rightarrow 0$ 
should be approached by fixing (in this case) two 
renormalized physical parameters which may be taken as the mass $M_RL_{\rm phy}$ 
and coupling constant $G_RL_{\rm phy}$ - expressed in units 
of the physical length $L_{\rm phy}$.
Perturbation theory allows us to relate these renormalized dimensionless
quantities to their bare lattice counterparts 
$$G_RL_{\rm phy}\sim gL,\quad M_R^2L_{\rm phy}^2\sim m^2L^2-Cg^2L^2\ln{\left(\mu a\right)}$$
where $\mu$ is the mass scale associated to the renormalization point
and $C$ is a numerical constant.
Along such an RG trajectory the value $\lambda=\frac{m}{2g}$
can then be related to its constant (continuum) value $\Lambda_R$ via
the relation
$$\Lambda_R\sim
\lambda\left(1-\frac{C}{\lambda^2}\ln{\left(\frac{1}{L}\right)}
\right)^{\frac{1}{2}}$$
For small enough $L<L_c$ the log term on the right is small and
$\lambda\sim\Lambda_R$. If this is the case the soliton action $S_{\rm DW}\sim
gL\lambda^3$ is approximately constant and the
corresponding free energy of such configurations {\it negative}
for any value of the bare parameters.
This would correspond to finite volume supersymmetry
breaking. Conversely for large enough $L$ the log term will dominate and
lead to an infinite soliton action as $L\to\infty$ for any value of
the bare parameters. In this limit the solitons should disappear and
supersymmetry should be restored. The cross-over between these
two behaviors occurs for
$$L\sim L_c=Ae^{-\lambda^2}$$
These conclusions are in agreement with the reasoning
presented 
in \cite{witten}. 

While this work was in preparation we received a preprint 
\cite{Beccaria}
in which the same model is studied in a Hamiltonian framework. 
The conclusions of this study are broadly in agreement with ours. 

\newpage
\section*{Appendix: the algorithm for determining 
the Pfaffian of an antisymmetric matrix.}

In this chapter we describe the algorithm for determining
the Pfaffian of an arbitrary antisymmetric $2N$ by $2N$ matrix $M$,
which is defined as follows:
\be
\Pf M = \frac 1{N!2^N}\ \varepsilon_{\a_1,\b_1,\ldots,\a_N,\b_N}
M_{\a_1,\b_1}\ldots M_{\a_N,\b_N}
\label{Pf}
\ee
The algorithm utilizes the following theorem.
\\
\\
{\it Theorem.} If $P$ is a matrix such that an antisymmetric matrix 
$M$ can be represented as $M=P^TJP$ where 
$J={\rm diag}(i\g_3,i\g_3, ... ,i\g_3)$ is a block-diagonal 
matrix then $\Pf M=\det P$ (here $C=i\g_3$ is the Euclidean 
representation of the charge conjugation matrix 
for two-dimensional system).
\\
\\
The theorem can be proved using the representation of Pfaffian
in terms of an integral over a Grassmann 2N-vector $\t$. 
Defining $\tt = P\t$ we have:
$$
\Pf M \equiv \int d\t e^{-\frac12 \t^TM\t}=
\int d\t e^{-\frac12 \t^T P^TJP \t}=
\int d\t e^{-\frac12 \tt^T J \tt}=
$$
$$
=\int d\t \frac 1{N!}[\tt_{2n-1} \tt_{2n}]^N=
\int d\t \frac 1{N!}[P_{2n-1,\a}\t_\a P_{2n,\b}\t_\b]^N=
$$
$$
=\int d\t P_{1,\a_1}\t_{\a_1} P_{2,\b_1}\t_{\b_1} \ldots
P_{2N-1,\a_N}\t_{\a_N} P_{2N,\b_N}\t_{\b_N}=
$$
$$
=\varepsilon_{\a_1,\b_1,\ldots,\a_N,\b_N}
P_{1,\a_1}P_{2,\b_1}\ldots P_{2N-1,\a_N}P_{2N,\b_N}=\det P
$$
\\
Notice that the matrix $P$ is not orthogonal ($P^T\ne P^{-1}$),
hence it is not associated with any basis transformation
in $2N$-dimensional vector space.

The above theorem can be given an alternative formulation.
Defining $Q=P^{-1}$ leads to the following statement:
if $Q^TMQ=J$ then $\Pf M=(\det Q)^{-1}$. This formulation is
used in the algorithm we describe below. The purpose of the algorithm
is to represent a given antisymmetric matrix $M$ in terms 
of a triangular matrix $Q$ so that the $\det Q$ and hence the 
$\Pf M$ can be found easily.
\\
\\
{\it The algorithm task.} Given an arbitrary antisymmetric $2N$ by $2N$ 
matrix $M$ find a triangular matrix $Q$ such that
$Q^TMQ=J$, $J={\rm diag}(i\g_3,i\g_3, ... ,i\g_3)$.
\\

The triangular matrix $Q=\{q_i\}$ represented in terms of its columns 
$q_i$ will satisfy the relation above {\it iff} its 
columns satisfy the following conditions:
$$
(q_{2i-1}Mq_{2j-1})=0,\ \ \ \ (q_{2i}Mq_{2j})=0
$$
$$
(q_{2i}Mq_{2j-1})=-(q_{2j-1}Mq_{2i})=\delta_{ij}
$$
The following algorithm 
by construction leads to such a matrix $Q$.
\\
\\
{\it The algorithm.}\\
1. Establish a unary $2N$ by $2N$ matrix 
$Q={\rm diag}\{1,1,...,1\}=\{e_i\}$, where unary vectors
$e_i$ are columns of $Q$, $i=1,2,...,2N$.\\
2. For odd values $i=1,3,...,2N-1$ repeat the following steps:\\
2-a. Leave $e_i$ as is.\\
2-b. Redefine $e_{i+1}\- e_{i+1}/(e_{i+1}Me_i)$\\
2-c. For $k=i+2,i+3,...,2N$ redefine
$e_k\- e_k-e_i(e_{i+1}Me_k)+e_{i+1}(e_iMe_k)$\\
\\
Notice that in 2-c the vector $e_{i+1}$ is used 
after it is redefined in 2-b.

\newpage
\section*{Acknowledgments}
Simon Catterall was supported in part by DOE grant DE-FG02-85ER40237.

\newpage
\section*{Figures.}

\end{document}